\documentclass[aps,preprint,groupedaddress]{revtex4}
\usepackage{amsmath} 
\usepackage{amssymb} 
\usepackage{amsfonts}
\usepackage[dvips]{graphicx} 
\usepackage[]{epsfig} 
\begin{document}
\bibliographystyle{apsrev} 

\title{Coupling hydrophobic, dispersion, and electrostatic contributions in continuum solvent models} 
\author{J. Dzubiella}
\email[e-mail address:] {jdzubiella@ucsd.edu} 
\affiliation{NSF Center for Theoretical Biological Physics (CTBP),} 
\affiliation{Department of
Chemistry and Biochemistry, University of California, San Diego, La
Jolla, California 92093-0365} 
\author{J.~M.~J. Swanson} 
\affiliation{NSF Center for Theoretical Biological Physics (CTBP),}
\affiliation{Department of Chemistry and Biochemistry, University of
California, San Diego, La Jolla, California 92093-0365}
\author{J.~A. McCammon} 
\affiliation{NSF Center for Theoretical Biological Physics (CTBP),} 
\affiliation{Department of Chemistry and
Biochemistry, University of California, San Diego, La Jolla,
California 92093-0365}

\date{\today}

\begin{abstract}
Recent studies of the hydration of micro- and nanoscale solutes have
demonstrated a strong {\it coupling} between hydrophobic, dispersion
and electrostatic contributions, a fact not accounted for in current
implicit solvent models. We present a theoretical formalism which
accounts for coupling by minimizing the Gibbs free energy with respect
to a solvent volume exclusion function. The solvent accessible surface
is output of our theory. Our method is illustrated with the hydration
of alkane-assembled solutes on different length scales, and captures
the strong sensitivity to the particular form of the solute-solvent
interactions in agreement with recent computer simulations.
\end{abstract}


\maketitle Much progress has been made in the last decade in the
understanding of hydrophobic solvation on different length scales
\cite{chandler:review,pratt2}. Most of this work has been devoted to
study solvation of purely repulsive, hard sphere-like solutes, while
less attention has been given to the influence and incorporation of
dispersion or electrostatic contributions.  Likewise, an entire field
in the biophysical community has explored electrostatic solvation
effects in the absence or uncoupled addition to hydrophobic
considerations \cite{roux:biochem}.  Recently, however, several
computer simulations have demonstrated a strong coupling between
hydrophobicity, solute-solvent dispersion attractions, and
electrostatics.  For example, a simulation of explicit water between
paraffin plates revealed that hydrophobic attraction and dewetting
phenomena are strongly sensitive to the nature of solute-solvent
dispersion interactions \cite{huang:jpcb}. Similarly, simulations of
hydrophobic channels \cite{dzubiella:channel1, vaitheesvaran} and
nanosolutes \cite{dzubiella:jcp:2003} have shown that charged solutes,
which attract the dipolar solvent due to increasing electric field
strength close to the solute surface, strongly affect the dewetting
behavior and potentials of mean force (pmf).  A fully atomistic
simulation of the folding of the two-domain protein BphC enzyme
\cite{zhou:science} further supported coupling by showing that the
region between the two domains was completely dewetted when
solvent-solute van der Waals (vdW) and electrostatic interactions were
turned off, but accommodated 30$\%$ of the density of bulk water with the
addition of vdW attractions, and 85-90$\%$ with the addition of
electrostatics, in accord with experimental results. Finally, Liu {\it
et al.}  recently observed a dewetting transition in the collapse of
the melittin tetramer which was strongly sensitive to the type and
location of the hydrophobic residues proving that these observations
apply to realistic biomolecular systems~\cite{berne:nature}.
 
In this letter we propose a continuum description of solvation that
explicitly couples hydrophobic, dispersion and electrostatic
contributions. Similar to the approach of Parker {\it et al.} in their
study of bubble formation at hydrophobic surfaces \cite{attard}, we
express the Gibbs free energy as a functional of the solute cavity
shape, the latter given by the volume exclusion function of the
solvent, and obtain the optimal shape by minimization. This leads to
an expression similar to the Laplace-Young equation for the
geometrical description of capillary surfaces \cite{kralchevsky}, but
in contrast to existing approaches {\it explicitly} includes the
inhomogeneous distributions of dispersion and electrostatic
contributions as well as curvature corrections.  Geometry-based
approaches similar to our formalism exist in related fields, such as
the Helfrich description of membranes shapes \cite{kralchevsky},
wetting in colloids and granular media \cite{kralchevsky}, and
electrowetting \cite{electrowetting}.  We stress that, as opposed to
other implicit solvent models \cite{roux:biochem}, the solvent
accessible surface (SAS) is an output of our theory.  This surface
encloses the optimal solvent accessible volume and should not be
confused with the canonical SAS \cite{roux:biochem} which is simply
the union of probe-inflated spheres. We begin by verifying that our
method is able to describe the solvation of small alkanes on molecular
scales. We then demonstrate that it captures the strong sensitivity of
dewetting and hydrophobic hydration to solute-solvent interactions on
larger scales for a model system of two alkane-assembled spheres. In
this striking example the strong hydrophobic attraction decreases
almost two orders of magnitude in units of the thermal energy, $k_BT$,
and dewetting is partially or completely suppressed when realistic
dispersion and electrostatic contributions are included. We expect our
approach to be particularly useful in solvation studies of proteins
where the hydrophobic surfaces are highly irregular and laced with
hydrophilic units, and a unified description of hydration on different
length scales is important
\cite{chandler:review,rossky:nature,berne:nature}.

Let us consider an assembly of solutes with arbitrary shape and
composition surrounded by a dielectric solvent in a macroscopic volume
$\cal W$.  We define a subvolume $\cal V$ empty of solvent for which
we can assign a volume exclusion function in space given by $v(\vec
r)= 0$ for $r \in \cal V$ and $v(\vec r)= 1$ elsewhere. We assume that
the surface bounding the volume is continuous and closed.  The
absolute volume $V$ and surface $S$ of $\cal V$ can then be expressed
as functionals of $v(\vec r)$ via $V[v]=\int_{\cal W}{\rm d}^3r
\;[1-v(\vec r)]$ and $S[v]=\int_{\cal W}{\rm d}^3r \;|\nabla v(\vec
r)|$, where $\nabla\equiv\nabla_{\vec r}$ is the usual gradient
operator. The density distribution of the solvent is given by
$\rho(v(\vec r))=\rho_0 v(\vec r)$, where $\rho_0$ is the bulk density
of the solvent at fixed temperature and pressure. The solutes' 
positions and conformations are fixed.

We suggest expressing the Gibbs free energy $ G[v]$ of the system as a
{\it functional} of $v(\vec r)$ and obtaining the optimal volume and
surface via minimization $\delta G[v]/\delta v[\vec r]=0$.  We adopt
the following ansatz for the Gibbs free energy:
\begin{eqnarray}
 G[v] &=&  P V[v]+\int_{\cal W}{\rm d}^3r\;\gamma(v)|\nabla
 v(\vec r)|+ \int_{\cal W}{\rm d}^3r\;\rho(v) U(\vec r)\nonumber \\ &+& \frac{\epsilon_0}{2}\int_{\cal W}{\rm
 d}^3r\;\{\nabla\Psi(v)\}^2\epsilon(v).
\label{eq:grand}
\end{eqnarray}
The first term in (\ref{eq:grand}) is the energy of creating a cavity
in the solvent against the difference in bulk pressure between the
liquid and vapor, $P=P_l-P_v$.  The second term describes the
energetic cost due to solvent rearrangement close to the cavity
surface in terms of a coefficient $\gamma$. This interfacial energy
penalty is thought to be the main driving force for hydrophobic
phenomena \cite{chandler:review}. The coefficient $\gamma$ is not only
a solvent specific quantity but also depends on the local topology of
the surface \cite{rossky:nature}, i.e., it is a function of the volume
exclusion function, $\gamma=\gamma(v(\vec r))$. The exact form of this
function is not known.  For planar macroscopic solvent-cavity
interfaces $\gamma$ is usually identified by the liquid-vapor surface
tension, $\gamma_{{\rm lv}}$, of the solvent, which we will also
employ here. In the following we make a {\it local curvature
approximation}, i.e. we assume that $\gamma$ can be expressed solely
as a function of the local mean curvature of the interface defined by
$v$, $\gamma(v(\vec r))=\gamma(H(\vec r))$, with $H(\vec
r)=(\kappa_1(\vec r)+\kappa_2(\vec r))/2$, where
$\kappa_1$ and $\kappa_2$ are the two principal curvatures. We then
apply the first order curvature correction to $\gamma$ given by
scaled-particle theory \cite{stilinger}, the commonly used ansatz to
study the solvation of hard spheres, arriving at
\begin{eqnarray}
\gamma(H({\vec r}))=\gamma_{{\rm lv}}(1+2\delta H(\vec r)),
\label{H}
\end{eqnarray}
where $\delta$ is a constant and positive length expected to be of the
order of the solvent particle size \cite{stilinger}. The curvature is
positive or negative for concave or convex surfaces, respectively.
Note that this leads to an increased surface tension for concave
surfaces, in agreement with the arguments of Nicholls {\it et
al.}~\cite{nicholls} in their study of alkanes.  It has been shown by
simulations that (\ref{H}) predicts the interfacial energy of growing
a spherical cavity in water rather well for radii $\gtrsim
3$\AA~\cite{huang:jpc}.

The third term in (\ref{eq:grand}) is the total energy of the
non-electrostatic solute-solvent interaction given a density
distribution $\rho(v)$. The energy $U(\vec r)=\sum_{i} U_i(\vec r)$ is the sum of
the short-ranged repulsive and long-ranged (attractive) dispersion
interactions $U_i$ between each solute atom $i$ and a solvent
molecule.  Classical solvation studies typically represent $U_i$ as an
isotropic Lennard-Jones (LJ) potential, $ U_{\rm
LJ}(r)=4\epsilon[(\sigma/r)^{12}-(\sigma/r)^6]$, with an energy scale
$\epsilon$ and a length scale $\sigma$.  The importance of treating
dispersion interactions independently as opposed to absorbing them in
to the surface tension term, has been emphasized by Gallicchio {\it et
al.} in their study of cyclic alkanes \cite{gallicchio:jpcb}.

The fourth term in (\ref{eq:grand}) describes the total energy of the
electrostatic field expressed by the local electrostatic potential
$\Psi(v(\vec r))$ and the position-dependent dielectric constant
$\epsilon(\vec r)$ assuming linear response of the dielectric solvent.
In general, the electrostatic potential $\Psi$ can be evaluated by
Poisson's equation, $ \nabla\cdot[\epsilon(\vec r)\nabla\Psi(\vec
r)]=-\lambda(\vec r)/\epsilon_0$, where $\lambda(\vec r)$ is the
solute's charge density distribution. The most common form for the
dielectric function, $\epsilon(\vec r)$, is proportional to the volume
exclusion function $v(\vec r)$ \cite{roux:biochem}
\begin{eqnarray}
\epsilon(v(\vec r))=\epsilon_v+v(\vec r)(\epsilon_l-\epsilon_v),
\label{e}
\end{eqnarray}
where $\epsilon_v$ and $\epsilon_l$ are the dielectric constants
inside and outside the volume $\cal V$, respectively.
 
Plugging in (\ref{H}) and (\ref{e}) in  functional (\ref{eq:grand})
and using the calculus of functional derivatives, the minimization
yields
\begin{eqnarray}
0 = P-2\gamma_{{\rm lv}}\left[H(\vec r)+\delta K(\vec r)\right]-\rho_0 U(\vec r)-\frac{\epsilon_0}{2}[\nabla\Psi(\vec r)\epsilon(\vec r)]^2\left(\frac{1}{\epsilon_l}-\frac{1}{\epsilon_v}\right).
\label{diff}
\end{eqnarray}
Eq. (\ref{diff}) is an ordinary second order differential equation for
the optimal solvent accessible volume and surface expressed in terms
of pressure, surface curvatures, dispersion interactions, and
electrostatics, all of which have dimensions of force per surface area
or energy density. $K(\vec r)=\kappa_1(\vec r)\kappa_2(\vec r)$ is the
Gaussian curvature and follows from the variation of the surface
integral over $H(\vec r)$ in (\ref{eq:grand}). Thus, in our approach
the geometry of the surface, expressed by $H$ and $K$, is directly
related to the inhomogeneous dispersion and electrostatic energy
contributions.  Note that the SAS is presently defined with respect to
the LJ centers of the solvent molecules.

In the following we illustrate solutions of (\ref{diff}) in spherical
and cylindrical symmetries.  For a spherical solute (\ref{diff})
reduces to a function of $R$, the radius of the solvent accessible
sphere, $H=-1/R$ and $K=1/R^2$.  In cylindrical symmetry the SAS
can be expressed by a one dimensional shape function $r(z)$, where $z$
is the coordinate on the symmetry axis in $z$-direction and $r$ the
radial distance to it. The three-dimensional surface is obtained by
revolving $r(z)$ around the symmetry axis. We express $r=r(t)$ and
$z=z(t)$ as functions of the parameter $t$.  The principal curvatures
are then given by $ \kappa_1=-z'/(r\sqrt{r'^2+z'^2})$ and
$\kappa_2=(z'r''-z''r')/((r'^2+z'^2)^{3/2}) $, where the primes
indicate the partial derivative with respect to $t$. We solve
(\ref{diff}) and Poisson's equation numerically, using standard
forward time relaxation schemes.

\begin{figure}[htb]
 \begin{center}
    \epsfig{file=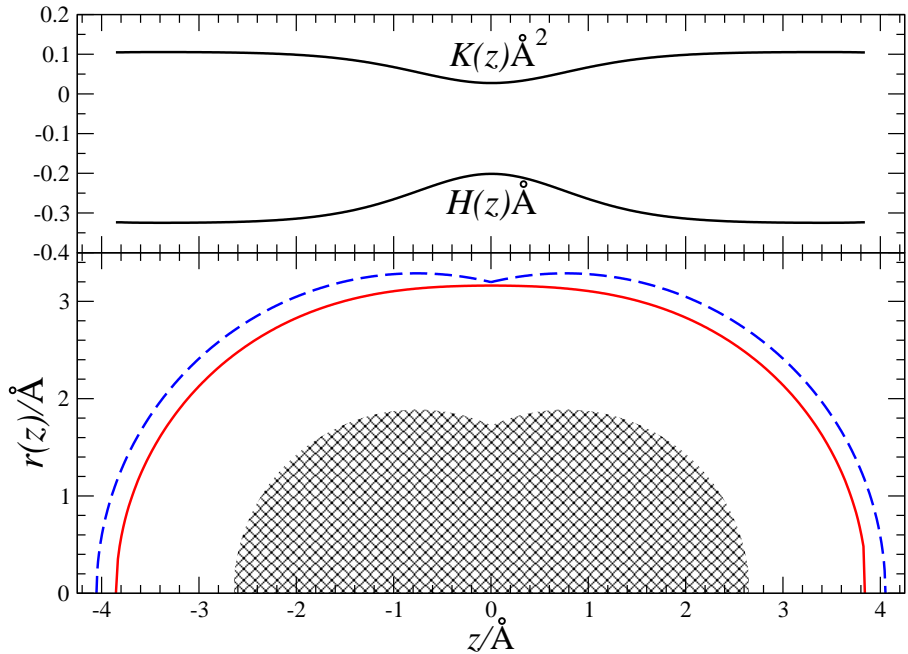, width=11cm, angle=-90}
   \caption{Mean $H(z)$ and Gaussian $K(z)$ curvature and shape
   function $r(z)$ (solid lines) for ethane. The canonical SAS (dashed
   line) from rolling a probe sphere with radius $r_p=1.4$\AA~ over
   the vdW surface (shaded region) is also shown.}
 \label{fig:ethane}
 \end{center}
 \end{figure}

We now study the solvation of methane and ethane in water and compare
our results to the SPC explicit water simulations by Ashbaugh {\it et
al.}  \cite{ashbaugh:biophys}, in which the alkanes are modeled
by neutral LJ spheres.  The LJ water-atom parameters are
$\epsilon=0.8941$kJ/mol and $\sigma=3.45$\AA~ for CH$_4$, and
$\epsilon=0.7503$kJ/mol and $\sigma=3.47$\AA~ for CH$_3$, and the bond
length of ethane is $1.53$\AA.  We fix the liquid-vapor surface
tension for SPC water at 300K to $\gamma_{\rm lv}=65$mJ/m$^2$
\cite{huang:jpc}. Since we deal with water under ambient conditions
the pressure term can be neglected and the length $\delta$ remains the
only free parameter. For methane we can reproduce the simulation
solvation energy with a fit $\delta=0.85$\AA.  This is in good
agreement with Huang {\it et al.}  \cite{huang:jpc} who measured
$\delta=0.76\pm0.05$\AA~ for SPC water. Solving the cylindrically
symmetric problem for the diatomic ethane with the same
$\delta=0.85$\AA, we obtain a fit-parameter-free $ G=11.48$kJ/mol,
which is only 7$\%$ larger than the simulation results. Alternatively,
the best fit $\delta=0.87$\AA~ reproduces the simulation energy
exactly.  This is surprisingly good agreement given the crude
curvature correction we apply and the fact that the large curvature of
the system varies locally in space. This supports the validity of our
continuum approach down to a molecular scale. The curvature and shape
functions $H(z)$, $K(z)$, and $r(z)$ are plotted in
Fig.~\ref{fig:ethane} together with the vdW surface and the canonical
SAS obtained from rolling a probe sphere with a typically chosen
radius $r_p=1.4$\AA~ over the vdW surface \cite{roux:biochem}. Away
from the center of mass $|z|\gtrsim 1$\AA~ the curvatures follow the
expected trends $H\simeq -1/R$ and $K\simeq 1/R^2$ with $R\simeq
3.1$\AA~ for the spherical surfaces. The surface resulting from our
theory is smaller than the canonical SAS, and is smooth at the center
of mass ($z=0$) where the canonical SAS has a kink. Thus our surface
has a smaller mean curvature at $z=0$ and an almost zero Gaussian
curvature, which is typical for a cylinder geometry for which one of
the principal curvatures is zero.  These results may justify the use
of smooth surfaces in coarse-grained models of closely-packed
hydrocarbon surfaces, a possibility we will now explore with
solvation on larger length scales where dewetting effects can occur.

Let us consider two spherical solutes which we assume to be
homogeneously assembled of CH$_2$ groups with a uniform density
$\rho$=0.024\AA$^{-3}$ up to a radius $R_0=15$\AA, defined by the maximal
distance between a CH$_2$ center and the center of the solute. The
CH$_2$-water LJ parameters are $\epsilon=0.5665$kJ/mol and
$\sigma=3.536$\AA. Similar ones have been used by Huang {\it et al.}
\cite{huang:jpcb} to study dewetting between paraffin plates. The
integration of the CH$_2$-water LJ interaction over the volume of a
sphere leads yields a 9-3 like potential for the interaction between
the center of the paraffin sphere and a water molecule
\cite{huang:jpcb:2002}.  The intrinsic, nonelectrostatic solute-solute
interaction $U_{\rm ss}(s)$ can be obtained in a similar fashion. The
solvation of the two solutes is studied for a fixed surface-to-surface
distance which we define as $s_0=r_{12}-2R_0$, where $r_{12}$ is
the solute center-to-center distance. We obtain an effective SAS radius of
one sphere of about $R \simeq R_0+2.4$\AA~ so that the effective
surface-to-surface distance is roughly $s \simeq s_0-4.8$\AA. Sine we
are also interested in the effects of charging up the solutes we place
opposite charges $\pm Ze$, where $e$ is the elementary charge, in the
center or on the edge of the two spheres.

In the following we focus on a separation distance of $s_0=8$\AA~ to
investigate the influence of different contributions to the energy
functional on the shape function, $r(z)$, and the curvatures, $K(z)$
and $H(z)$. For $s_0 = 8$\AA, it follows that $s\simeq 3.2 $\AA, such
that two water molecules could fit between the solutes on the
$z$-axis. We systematically change the solute-solute and
solute-solvent interactions, as summarized in Tab.~I. We begin with
only the LJ repulsive interactions in system I and then add a
curvature correction with $\delta=0.75$\AA, vdW attractions, and
sphere-centered charges $Z = 4$ and $Z = 5$ in systems II-V,
respectively.  To study the influence of charge location, we shift
each charge to the edge of the spheres such that they are $8$\AA~apart
and reduce their magnitude to $Z =1$ (system VI).  The surface tension
and dielectric constant of the vapor and liquid are fixed to
$\gamma_{\rm lv}=72$mJ/m$^2$, $\epsilon_v=1$, and $\epsilon_l=78$,
respectively.
\begin{table}
\begin{center}
\begin{tabular}{l | c c c c c}
  System & $\delta/{\rm \AA}\;\;\;$ & vdW attraction & $\;\;\;Z$ & $W(s_0)/k_BT$ & dewetted\\ 
\hline 
I   & 0.00 & no & 0 & -57.6 &yes \\ 
II  & 0.75 & no & 0 & -34.1 &yes \\ 
III & 0.75 & yes & 0 & -6.3 &yes \\ 
IV  & 0.75 & yes & 4 & -9.2 &yes \\ 
V   & 0.75 & yes & 5 & -5.1 &no \\ 
VI  & 0.75 & yes & 1 (oc)& -1.3 &no \\
\end{tabular}
\caption{Studied systems for two alkane-assembled spherical
solutes. $W(s_0)$ is the inter-solute pmf. If $r(z=0)\neq0$ the system
is 'dewetted'. In system VI the solutes' charge is located off-center
(oc) at the solute surface.}
\label{tab4}
\end{center}
\end{table}

 \begin{figure}[htb]
 \begin{center}
    \epsfig{file=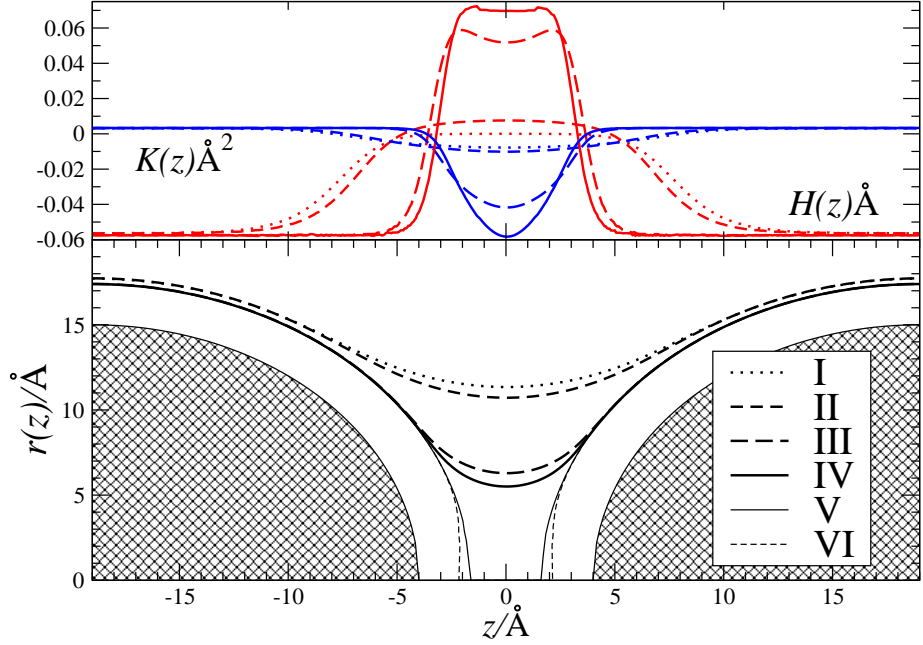, width=11cm, angle=-90}
   \caption{Mean $H(z)$ and Gaussian $K(z)$ curvatures and shape
   function $r(z)$ for two alkane-assembled solutes of radius
   $R_0=15$\AA~ (shaded region) for systems I-VI. Curvatures are not
   shown for the 'wet' systems V and VI. }
 \label{fig:nano}
 \end{center}
 \end{figure}

The results for the curvatures and SAS, defined by $r(z)$, for systems
I-VI are shown in Fig.~\ref{fig:nano}. Away from the center of mass
($|z|\gtrsim 10$\AA) systems I-VI show very little difference. The
curvatures are $H\simeq -1/R$ and $K\simeq 1/R^2$ with $R \simeq
17.4$\AA. Close to the center of mass ($z\simeq 0$), however, the
influence of changing the parameters is considerable. In system I,
Eq.~(\ref{diff}) reduces to the minimum surface equation $H(z)=0$ for
$z\simeq 0$. For two adjacent spheres the solution of this equation is
the catenoid $r(z)\simeq{\rm cosh}(z)$, which features zero mean
curvature ($\kappa_1$ and $\kappa_2$ cancel each other) and negative
Gaussian curvature. This leads to a vapor bubble bridging the solutes.
When curvature correction is applied (system II) the mean curvature
becomes nonzero and positive (concave) at $z\simeq 0$, while the
Gaussian curvature grows slightly more negative. As a consequence the
total enveloping surface area becomes larger and the solvent
inaccessible volume shrinks, i.e. the value of the shape function at
$z \simeq 0$ decreases. Turning on solute-solvent dispersion
attraction amplifies this trend significantly as demonstrated by
system III. Mean and Gaussian curvatures increase fivefold, showing
strongly enhanced concavity, and the volume empty of water decreases
considerably, expressed by $r(z=0)\simeq 10.7$\AA~ dropping to
$r(z=0)\simeq 6.3$\AA. These trends continue with the addition of
electrostatics in system IV.  When the sphere charges are further
increased from $Z=4$ to $Z=5$ (system IV$\rightarrow$V), we observe a
wetting transition: the bubble ruptures and the SAS jumps to the
solution for two isolated solutes, where $r(z\simeq 0)=0$. The same
holds when going from III to VI, when only one charge, $Z=1$, is
placed at each of the solutes' surfaces. Importantly, this
demonstrates that the present formalism captures the sensitivity of
dewetting phenomena to specific solvent-solute interactions as
demonstrated in previous studies
\cite{huang:jpcb,dzubiella:channel1,vaitheesvaran,dzubiella:jcp:2003,zhou:science,berne:nature}.
Note that the SAS at $|z|\simeq\pm2$\AA~ is closer to the solutes in
VI compared to V due to the proximity of the charge to the
interface. Clearly, the observed effects, in particular the transition
from III to VI, cannot be described by existing solvation models,
which use the SAS \cite{roux:biochem}, or effective surface tensions
and macroscopic solvent-solute contact angles \cite{attard} as input.

The significant change of the SAS with the solute-solvent interaction
has a strong impact on the pmf, $W(s_0)= G(s_0)- G(\infty)+U_{\rm
ss}(s_0)$. Values of $W(s_0=8{\rm\AA})$ are given in Tab.~I. From
system I to VI the total attraction between the solutes decreases
almost two orders of magnitude. Interestingly, the curvature
correction (I$\rightarrow$II) lowers $W$ by a large 23.5$k_BT$, even
though $R\gg\delta$. A striking effect occurs when vdW contributions
are introduced (II$\rightarrow$III): the inter solute attraction
decreases by $\simeq 28k_BT$ while the dispersion solute-solute
potential, $U_{\rm ss}(s_0=8{\rm \AA})$, changes by only
-0.44$k_BT$. Similarly, adding charges of $Z=5$ (III $\rightarrow$ V)
at the solutes' centers or $Z=1$ (III $\rightarrow$ VI) at the
solutes' surfaces decreases the total attraction by 1.2$k_BT$ and
6k$_B$T, respectively. Note that the total attraction decreases
although electrostatic attraction has been added between the
solutes. The same trends have been observed in explicit water
simulations of a similar system of charged hydrophobic nanosolutes~\cite{dzubiella:jcp:2003}.

These results clearly demonstrate that
solvation effects and solvent mediated phenomena are not only strongly
influenced by solute-solvent interactions, but that these interactions
are inherently coupled.  By including coupling, our formalism
captures the balance between hydrophobic, dispersive and electrostatic
forces which has been observed in previous studies
\cite{huang:jpcb,dzubiella:channel1,vaitheesvaran,dzubiella:jcp:2003,zhou:science,berne:nature}
but never described in a single theoretical framework.  Nonpolar and
polar coupling is expected to be crucial for a complete
characterization of biomolecular solvation.  The present formalism is
only limited by the crude curvature and dielectric descriptions
currently employed. Future efforts to improve these approximations
will be critical to accurately describe solvation effects on multiple
length scales and for more complicated geometries.

The authors thank Tushar Jain, John Mongan, and Cameron Mura for
useful discussions.  J.D. acknowledges financial support from a DFG
Forschungsstipendium. Work in the McCammon group is supported by NSF,
NIH, HHMI, CTBP, NBCR, and Accelrys, Inc.


\begin{thebibliography}{19}
\expandafter\ifx\csname natexlab\endcsname\relax\def\natexlab#1{#1}\fi
\expandafter\ifx\csname bibnamefont\endcsname\relax
  \def\bibnamefont#1{#1}\fi
\expandafter\ifx\csname bibfnamefont\endcsname\relax
  \def\bibfnamefont#1{#1}\fi
\expandafter\ifx\csname citenamefont\endcsname\relax
  \def\citenamefont#1{#1}\fi
\expandafter\ifx\csname url\endcsname\relax
  \def\url#1{\texttt{#1}}\fi
\expandafter\ifx\csname urlprefix\endcsname\relax\def\urlprefix{URL }\fi
\providecommand{\bibinfo}[2]{#2}
\providecommand{\eprint}[2][]{\url{#2}}

\bibitem[{\citenamefont{Chandler}(2005)}]{chandler:review}
\bibinfo{author}{\bibfnamefont{D.}~\bibnamefont{Chandler}},
  \bibinfo{journal}{Nature}  (\bibinfo{year}{2005}), \bibinfo{note}{in press,
  http://gold.cchem.berkeley.edu:8080/Pubs/DC202.pdf}.

\bibitem[{\citenamefont{Hummer et~al.}(2000)\citenamefont{Hummer, Garde,
  Garcia, and Pratt}}]{pratt2}
\bibinfo{author}{\bibfnamefont{G.}~\bibnamefont{Hummer et al.}},
  \bibinfo{journal}{Chem. Phys.} \textbf{\bibinfo{volume}{258}},
  \bibinfo{pages}{349} (\bibinfo{year}{2000}).

\bibitem[{\citenamefont{Roux}(1999)}]{roux:biochem}
\bibinfo{author}{\bibfnamefont{B.}~\bibnamefont{Roux}},
  \bibinfo{journal}{Biophys. Chem.} \textbf{\bibinfo{volume}{78}},
  \bibinfo{pages}{1} (\bibinfo{year}{1999}).

\bibitem[{\citenamefont{Huang et~al.}(2005)\citenamefont{Huang, Zhou, and
  Berne}}]{huang:jpcb}
\bibinfo{author}{\bibfnamefont{X.}~\bibnamefont{Huang et al.}},
  \bibinfo{journal}{J. Phys. Chem. B} \textbf{\bibinfo{volume}{109}},
  \bibinfo{pages}{3546} (\bibinfo{year}{2005}).

\bibitem[{\citenamefont{Dzubiella and Hansen}(2003)}]{dzubiella:channel1}
\bibinfo{author}{\bibfnamefont{J.}~\bibnamefont{Dzubiella}} \bibnamefont{and}
  \bibinfo{author}{\bibfnamefont{J.-P.} \bibnamefont{Hansen}},
  \bibinfo{journal}{J. Chem. Phys.} \textbf{\bibinfo{volume}{120}},
  \bibinfo{pages}{5001} (\bibinfo{year}{2003}).

\bibitem[{\citenamefont{Vaitheesvaran et~al.}(2004)\citenamefont{Vaitheesvaran,
  Rasaiah, and Hummer}}]{vaitheesvaran}
\bibinfo{author}{\bibfnamefont{S.}~\bibnamefont{Vaitheesvaran et al.}},
  \bibinfo{journal}{J. Chem. Phys.} \textbf{\bibinfo{volume}{121}},
  \bibinfo{pages}{7955} (\bibinfo{year}{2004}).

\bibitem[{\citenamefont{Dzubiella and Hansen}(2004)}]{dzubiella:jcp:2003}
\bibinfo{author}{\bibfnamefont{J.}~\bibnamefont{Dzubiella}} \bibnamefont{and}
  \bibinfo{author}{\bibfnamefont{J.-P.} \bibnamefont{Hansen}},
  \bibinfo{journal}{J. Chem. Phys.} \textbf{\bibinfo{volume}{119}},
  \bibinfo{pages}{12049} (\bibinfo{year}{2004}).

\bibitem[{\citenamefont{Zhou et~al.}(2004)\citenamefont{Zhou, Huang, Margulis,
  and Berne}}]{zhou:science}
\bibinfo{author}{\bibfnamefont{R.}~\bibnamefont{Zhou et al.}},
  \bibinfo{journal}{Science} \textbf{\bibinfo{volume}{305}},
  \bibinfo{pages}{1605} (\bibinfo{year}{2004}).

\bibitem[{\citenamefont{Liu et~al.}(2005)\citenamefont{Liu, Huang, Zhou, and
  Berne}}]{berne:nature}
\bibinfo{author}{\bibfnamefont{P.}~\bibnamefont{Liu et al.}},
  \bibinfo{journal}{Nature} \textbf{\bibinfo{volume}{437}},
  \bibinfo{pages}{159} (\bibinfo{year}{2005}).

\bibitem[{\citenamefont{Parker et~al.}(1994)\citenamefont{Parker, Claesson, and
  Attard}}]{attard}
\bibinfo{author}{\bibfnamefont{J.~L.} \bibnamefont{Parker et al.}},
  \bibinfo{journal}{J. Phys. Chem.} \textbf{\bibinfo{volume}{98}},
  \bibinfo{pages}{8468} (\bibinfo{year}{1994}).

\bibitem[{\citenamefont{Kralchevsky and Nagayama}(Amsterdam)}]{kralchevsky}
\bibinfo{author}{\bibfnamefont{P.}~\bibnamefont{Kralchevsky}} \bibnamefont{and}
  \bibinfo{author}{\bibfnamefont{K.}~\bibnamefont{Nagayama}},
  \emph{\bibinfo{title}{Particles at Fluid Interfaces and Membranes}}
  (\bibinfo{publisher}{Elsevier}, \bibinfo{address}{2001},
  \bibinfo{year}{Amsterdam}).

\bibitem[{\citenamefont{Chou}(2001)}]{electrowetting}
\bibinfo{author}{\bibfnamefont{T.}~\bibnamefont{Chou}}, \bibinfo{journal}{Phys.
  Rev. Lett.} \textbf{\bibinfo{volume}{87}}, \bibinfo{pages}{106101}
  (\bibinfo{year}{2001}).

\bibitem[{\citenamefont{Cheng and Rossky}(1998)}]{rossky:nature}
\bibinfo{author}{\bibfnamefont{Y.-K.} \bibnamefont{Cheng}} \bibnamefont{and}
  \bibinfo{author}{\bibfnamefont{P.~J.} \bibnamefont{Rossky}},
  \bibinfo{journal}{Nature} \textbf{\bibinfo{volume}{392}},
  \bibinfo{pages}{696} (\bibinfo{year}{1998}).

\bibitem[{\citenamefont{Stillinger}(1973)}]{stilinger}
\bibinfo{author}{\bibfnamefont{F.~H.} \bibnamefont{Stillinger}},
  \bibinfo{journal}{J. Solution Chem.} \textbf{\bibinfo{volume}{2}},
  \bibinfo{pages}{141} (\bibinfo{year}{1973}).

\bibitem[{\citenamefont{Nicholls et~al.}(1991)\citenamefont{Nicholls, Sharp,
  and Honig}}]{nicholls}
\bibinfo{author}{\bibfnamefont{A.}~\bibnamefont{Nicholls et al.}},
  \bibinfo{journal}{Proteins} \textbf{\bibinfo{volume}{11}},
  \bibinfo{pages}{281} (\bibinfo{year}{1991}).

\bibitem[{\citenamefont{Huang et~al.}(2001)\citenamefont{Huang, Geissler, and
  Chandler}}]{huang:jpc}
\bibinfo{author}{\bibfnamefont{D.~M.} \bibnamefont{Huang et al.}},
  \bibinfo{journal}{J. Phys. Chem. B} \textbf{\bibinfo{volume}{105}},
  \bibinfo{pages}{6704} (\bibinfo{year}{2001}).

\bibitem[{\citenamefont{Gallicchio et~al.}(2000)\citenamefont{Gallicchio, Kubo,
  and Levy}}]{gallicchio:jpcb}
\bibinfo{author}{\bibfnamefont{E.}~\bibnamefont{Gallicchio et al.}},
  \bibinfo{journal}{J. Phys. Chem. B} \textbf{\bibinfo{volume}{104}},
  \bibinfo{pages}{6271} (\bibinfo{year}{2000}).

\bibitem[{\citenamefont{Ashbaugh et~al.}(1998)\citenamefont{Ashbaugh, Kaler,
  and Paulaitis}}]{ashbaugh:biophys}
\bibinfo{author}{\bibfnamefont{H.~S.} \bibnamefont{Ashbaugh et al.}},
  \bibinfo{journal}{Biophys. J.} \textbf{\bibinfo{volume}{75}},
  \bibinfo{pages}{755} (\bibinfo{year}{1998}).

\bibitem[{\citenamefont{Huang and Chandler}(2002)}]{huang:jpcb:2002}
\bibinfo{author}{\bibfnamefont{D.~M.} \bibnamefont{Huang}} \bibnamefont{and}
  \bibinfo{author}{\bibfnamefont{D.}~\bibnamefont{Chandler}},
  \bibinfo{journal}{J. Phys. Chem. B} \textbf{\bibinfo{volume}{106}},
  \bibinfo{pages}{2047} (\bibinfo{year}{2002}).

\end{thebibliography}
\end{document}